\documentclass[11pt]{article}
\usepackage{hyperref}
\pdfoutput=1
\begin{document}
\title{Bubble Oscillations and Motion under Vibration}
\author{T. J. O'Hern, B. Shelden, and J. R. Torczynski \\
\\\vspace{6pt} Engineering Sciences Center \\ Sandia National Laboratories \\ Albuquerque, NM 87185, USA}
\maketitle
%

Bubbles under vibration can behave in unusual ways, for example, moving downward against the force of buoyancy [1-4]. Fluid dynamics videos 
have been recorded to demonstrate some of the phenomena that occur when a gas-liquid interface is forced to break up by vertical vibrations 
within critical ranges of frequency and amplitude and when the bubbles under the disturbed interface subsequently interact. While the bubble 
downward motion due to the Bjerknes force is well known at acoustic frequencies close to the bubble resonant frequency, these experiments, 
like those of [3], demonstrate that these effects can be observed at relatively low frequencies as well.  \\

Experiments were performed  in a thin, quasi-two-dimensional rectangular acrylic box (0.6 cm thick and 7.6 cm wide) partially filled with 
20-cSt polydimethylsiloxane (PDMS) silicone oil with an overlying layer of ambient air. The test apparatus was subjected to sinusoidal axial 
vibration conditions that produced breakup of the gas-liquid free surface, producing liquid jets into the air, droplets pinching off from 
these jets, gas cavities in the liquid from impacts of these droplets, and bubble transport below the interface. Experiments were performed 
for frequencies of 100-300 Hz, with displacement amplitudes up to 400 microns and vibrational accelerations up to 30 times the gravitational 
acceleration. The vibration conditions for the attached videos are 280 Hz frequency, 15 g acceleration, and 94 micron peak-to-peak displacement. Behaviors shown in the videos include the following.

\begin{enumerate}
\item Free surface breakup into jets and droplets, and formation of bubbles under the free surface.
\item Bubbles thus generated moving downward in the cell. 
\item Bubbles attracted to the first bubble deep in the cell and eventually merging to form a large bubble at the base of the cell. 
\item Bubble cluster at the base of the cell merging to form a larger bubble, which stabilizes at a levitated location below the free surface and acts to damp out the surface breakup. 
\item The levitated bubble interface and its breakup are similar to the free surface breakup into jets and droplets, but the jets in the bubble are facing downward.
\end{enumerate}
High-speed videos were recorded using Phantom V. 9.1 cameras acquiring up to 1000 frames per second. The observed phenomena are not unique to the particular PDMS oil used here, but have been observed with other oils over a broad range of viscosities, as well as in water. Applications include liquid fuel rockets, inertial sensing devices, moving vehicles, mixing processes, and acoustic excitation. 
\\

The videos showing the surface deformation and breakup and the bubble behavior can be seen in the ancillary files. This video was shown at the 
American Physical Society Division of Fluid Dynamics (APS DFD) Gallery of Fluid Motion 2011. The Gallery of Fluid Motion is an annual showcase 
of fluid dynamics videos and posters.
\\

Sandia National Laboratories is a multi-program laboratory managed and operated by Sandia Corporation, a wholly owned subsidiary of Lockheed 
Martin Corporation, for the U.S. Department of Energy's National Nuclear Security Administration under contract DE-AC04-94AL85000.

\footnotetext[1] {Brennen, C. E., Fundamentals of Multiphase Flows, Cambridge University Press, pp. 95-96, 2005.}

\footnotetext[2] {Jameson, G. J., The motion of a bubble in a vertically oscillating viscous liquid, Chemical Engineering Science, Vol. 21, pp. 35-38, 1966.}

\footnotetext[3] {Ellenberger, J., and Krishna, R., Levitation of air bubbles in liquid under low frequency vibration excitation, Chemical Engineering Journal, Vol. 62, pp. 5669-5673, 2007.}

\footnotetext[4] {Hashimoto, H., and Sudo, S., Surface Disintegration and Bubble Formation in Vertically Vibrated Liquid Column, AIAA Journal, Vol. 18, No. 4, pp. 442-449, 1980.}
\end{document}